\documentclass[preprint,journal]{vgtc}            

\usepackage{paralist}
\usepackage{balance}
\usepackage[svgnames]{xcolor}
\usepackage{tcolorbox}
\usepackage{tikz}
\usepackage{multirow}
\usepackage{fixltx2e}
\usepackage{wrapfig}

\newcommand*\circled[1]{\tikz[baseline=(char.base)]{
            \node[shape=circle,draw=white!50!black,fill=gray!15!white,inner sep=0.2pt, line width=0.5pt, text=black, font=\footnotesize] (char) {#1};}}
\newcommand*\teasercircled[1]{\tikz[baseline=(char.base)]{
            \node[shape=circle,draw=white!50!black,fill=gray!15!white,inner sep=0.4pt, line width=0.2pt, text=black, font=\small] (char) {#1};}}
\newcommand*\darkcircled[1]{\tikz[baseline=(char.base)]{
            \node[shape=circle,draw=blue!10!gray,fill=blue!10!gray,inner sep=0.5pt, line width=0.5pt, text=white, font=\footnotesize] (char) {#1};}}

\newcommand*\casebox[1]{\tikz[baseline=(char.base)]{
            \node[shape=rectangle,draw=white!30!black,fill=gray!15!white,inner sep=1pt, line width=0.5pt, text=black, font=\footnotesize] (char) {#1};}}

\newcommand{\audio}{\textcolor[RGB]{125, 87, 193}}
\newcommand{\text}{\textcolor[RGB]{63, 167, 250}}
\newcommand{\framekk}{\textcolor[RGB]{27, 178, 177}}

\onlineid{1223}

\vgtccategory{Research}

\title{LiveRetro: Visual Analytics for Strategic Retrospect in Livestream E-Commerce}

\author{%
\authororcid{Yuchen Wu}{0009-0005-8333-4405}, Yuansong Xu, Shenghan Gao, \authororcid{Xingbo Wang}{0000-0001-5693-1128}, Wenkai Song, Zhiheng Nie,\\ \authororcid{Xiaomeng Fan}{0000-0002-6302-8786}, and \authororcid{Quan Li}{0000-0003-2249-0728}
}

\authorfooter{
  \item Y. Wu, Y. Xu, S. Gao, and Q. Li are with the School of Information Science and Technology, ShanghaiTech University, and Shanghai Engineering Research Center of Intelligent Vision and Imaging, China. W. Song and X. Fan are with the School of Entrepreneurship and Management, ShanghaiTech University. Q. Li is the corresponding author. E-mail: wuych3,xuys1,gaoshh1,liquan,songwk,fanxm@shanghaitech.edu.cn.
  \item X. Wang is with Weill Cornell Medical College, Cornell University. E-mail: xingbo.wang@\{med.cornell.edu, connect.ust.hk\}.
  \item Z. Nie is with Be Friends Holding Limited. E-mail: 815194445@qq.com.
}

\abstract{Livestream e-commerce integrates live streaming and online shopping, allowing viewers to make purchases while watching. However, effective marketing strategies remain a challenge due to limited empirical research and subjective biases from the absence of quantitative data. Current tools fail to capture the interdependence between live performances and feedback. This study identified computational features, formulated design requirements, and developed \textit{LiveRetro}, an interactive visual analytics system. It enables comprehensive retrospective analysis of livestream e-commerce for streamers, viewers, and merchandise. \textit{LiveRetro} employs enhanced visualization and time-series forecasting models to align performance features and feedback, identifying influences at channel, merchandise, feature, and segment levels. Through case studies and expert interviews, the system provides deep insights into the relationship between live performance and streaming statistics, enabling efficient strategic analysis from multiple perspectives.
}


\teaser{
  \centering
  \includegraphics[width=\linewidth]{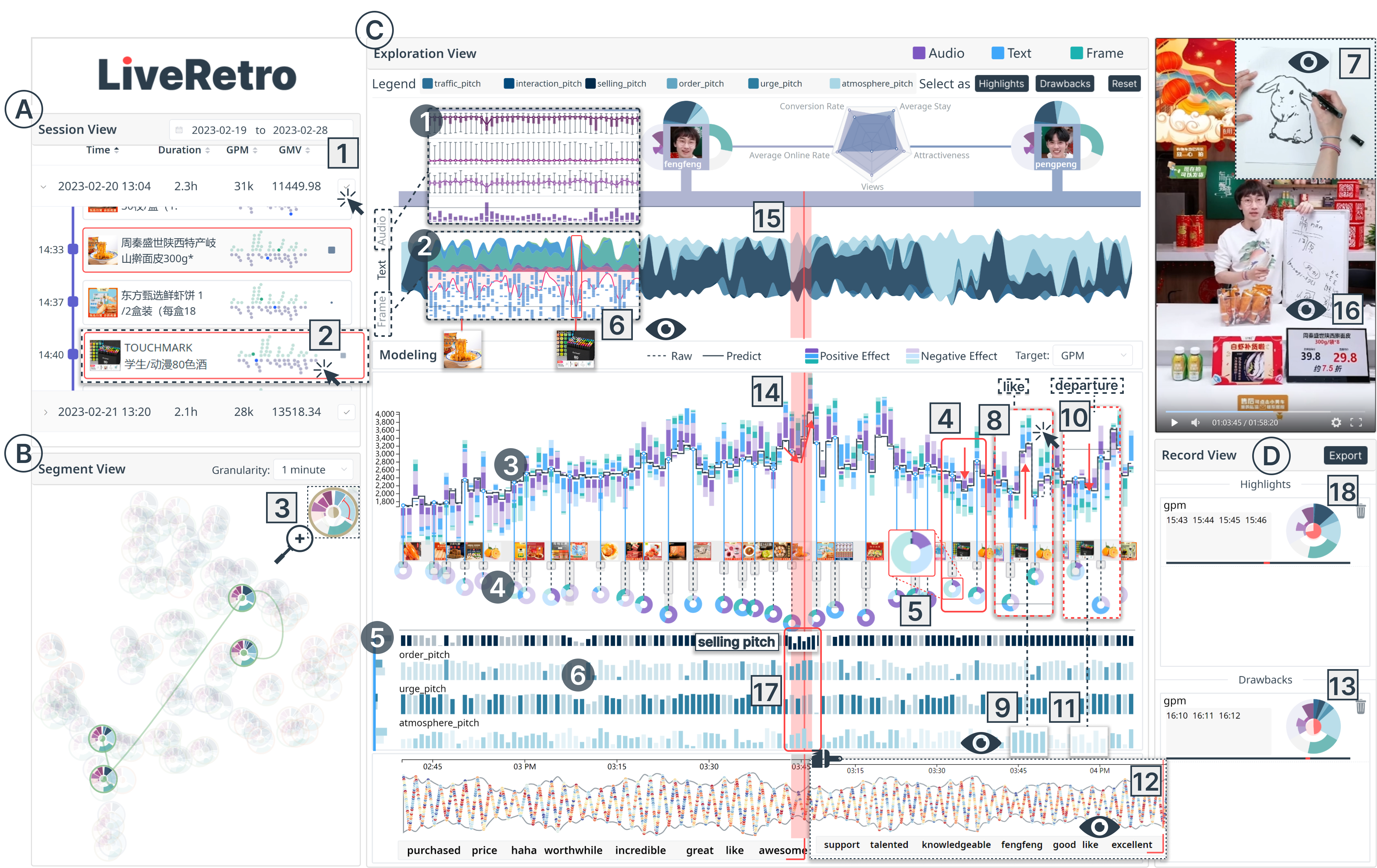}
  \vspace{-6mm}
  \caption{In Case I, the streamer's strategic analysis comprises: \protect\casebox{1} Select the session with the highest transaction amount per thousand views (GPM). \protect\casebox{2} Identify poor sales performance of the ``Marker Pen'' in the \protect\teasercircled{A} \textit{Session View}. \protect\casebox{3} Discover that the \textit{atmosphere pitch} dominates the \textit{Text} channel during that period in the\protect\teasercircled{B} \textit{Segment View}. \protect\casebox{4} Observe a low \textit{GPM} in the \protect\teasercircled{C} \textit{Exploration View}. \protect\casebox{5} Observe negative contributions from the \textit{Text} and \textit{Frame} channels in the merchandise-level summary. \protect\casebox{6} Find a scarcity of facial expressions and a high number of \textit{close-up}s in the \textit{Performance Inspection}. \protect\casebox{7} Observe in the live replay that the streamer was drawing a small rabbit on paper. \protect\casebox{8} Observe a peak in the \textit{Like} value during that time period. \protect\casebox{9} Discover the positive contribution of the \textit{atmosphere pitch} to the target. \protect\casebox{10} Observe low \textit{Departure} and \protect\casebox{11} negative contribution. \protect\casebox{12} Find an abundance of semantically similar comments in the \textit{Comment Reference} part. \protect\casebox{13} Draw the conclusion that streamers should maintain a balance in the utilization of various sales pitches, which is appended as \textit{Drawbacks} in the \protect\teasercircled{D} \textit{Record View}. \protect\casebox{14} Identify a period where the \textit{GPM} initially decreased but quickly rebounded. \protect\casebox{15} Observe a transition from the \textit{atmosphere pitch} to the \textit{selling pitch} and \textit{order pitch} during this period. \protect\casebox{16} Review video content revealing the streamer providing historical tidbits, and \protect\casebox{17} all three sales pitches found to have a positive impact. \protect\casebox{18} Summarize the combination of these three sales pitches as a wise choice and added as \textit{Highlights} to the \textit{Record View}.}
  \label{fig:teaser}
}

\graphicspath{{figs/}{figures/}{pictures/}{images/}{./}} 

\usepackage{tabu}                      
\usepackage{booktabs}                  
\usepackage{lipsum}                    
\usepackage{mwe}                       

\usepackage{mathptmx}                  

\begin{document}



\maketitle

\section{Introduction} 
\par Livestream e-commerce has emerged as a novel channel that blends entertainment with instant purchases by integrating live streaming with e-commerce, enabling viewers to make purchases while they watch. This presents a significant opportunity for value creation for retailers, brands, and digital platforms. However, the pressing challenge is how to effectively market in livestream e-commerce. Empirical studies have demonstrated that marketing strategy plays a critical role in the conceptual core of marketing practice in this domain~\cite{Morgan2018ResearchIM, wu2021forming,ram2019live}. An example of a marketing strategy in the context of live e-commerce is to limit the time of sale and the number of units available. However, achieving successful marketing outcomes entails more than merely employing plausible strategies. Issues often arise during the execution phase, where ``\textit{seemingly viable strategic plans fail to effect change}''~\cite{Nutt1983ImplementationAF}. Such problems emanate from various realities, including the organization's distinctive execution capabilities and constraints~\cite{Bonoma1985TheME}, genuine core competencies~\cite{hamel1990strategic}, and evolving internal operations of the organization, among others~\cite{hutt1988tracing,piercy2016market}. In essence, the dichotomy between marketing strategy development and implementation is often the underlying cause of many execution challenges~\cite{Cespedes1996ImplementingMS, haenlein2020navigating}. As a result, a comprehensive strategic retrospect is necessary to examine actual strategy execution and devise tailored optimizations for further exploitation.

\par Previous studies on livestream e-commerce have endeavored to illustrate the motivations for usage among streamers and consumers~\cite{Cai2019LiveSC,hou2020factors}, streamer characteristics~\cite{zhang2022characteristics,Yang2023ImpactOS}, and the factors that influence both usage intention and purchase intention~\cite{lu2021live,xu2020drives}. Some studies have also examined marketing strategies in livestream e-commerce from the perspective of sellers~\cite{wongkitrungrueng2020live,guo2022way}. However, most of these studies have primarily focused on developing theoretical models and providing implications as general guidelines, rather than providing a holistic and detailed assessment of strategic operations. Additionally, many of these studies have employed empirical investigations (e.g., online questionnaire) to examine perceptions and validate hypotheses, easily influenced by subjectivity and lead to biases~\cite{nickerson1998confirmation,ariely2003coherent,markus1987toward}
. To aid in comprehending the implications of the practical execution of live e-commerce strategy and to facilitate strategic retrospectives, commercially available analytical tools\footnote{\url{https://zhitou.feigua.cn}; \url{https://www.changuanjia.com}} are developed and subscribed to by professional streamers. However, these tools only provide rudimentary presentations of live replay and statistics, e.g., raw video and multi-line charts. Several studies have established that in live streaming and retail marketing, streamers/salespeople have a significant impact on viewers/customers~\cite{gauri2021evolution,li2021attachment,lin2021happiness,meng2021impact}. This indicates that, as the focus of strategy execution, the live performance of streamers has a profound influence on streaming statistics~\cite{wohn2018explaining,lu2018you}. Nonetheless, none of these tools explicate the connection between live performance and streaming statistics in an intuitive manner, leaving streamers with a ``click one, watch one'' back-and-forth approach to retrospective analysis. Thus, a systematic analytical approach is required to decode the relationship between live performance and streaming statistics to conduct strategic reviews and effectively evaluate actual marketing strategy implementation.

\par Visual analytics are effective approaches for analyzing video and time-series data. However, it remains challenging to conduct comprehensive strategic retrospect in livestream e-commerce.
First, \textbf{the characteristics of livestream e-commerce data make it challenging to process}. 
The need for multi-modal analysis naturally arises from a comprehensive analysis of livestream playback. Unlike previous studies focusing on presentation videos~\cite{wu2018multimodal, zeng2019emoco, zeng2022gesturelens, maher2021ffective} or user behavior videos~\cite{soure2021coux, batch2023uxsense,wong2023anchorage}, a single session of livestream commerce can last for several hours without interruption. In addition, streaming statistics encompass multiple metrics such as viewer count and order volume, rendering the data high-dimensional and prolonged. 
Second, \textbf{existing approaches cannot handle multi-modal video analysis and synchronous feedback simultaneously.} Previous studies have centered on examining information present within the video, including speech~\cite{yuan2019speechlens}, emotion~\cite{zeng2019emoco,zeng2020emotioncues}, gesture~\cite{zeng2022gesturelens}, and intra-modal correlation~\cite{wu2018multimodal,wang2021m2lens}. Nevertheless, livestream e-commerce involves a process where streamers deliver information, and viewers receive information while simultaneously responding through activities such as making purchases and joining the live stream. This specificity of the livestream context renders existing systems inadequate to meet the demands of joint analysis of live performances and feedback. 
Third, \textbf{multi-modal features and time-series feedback have distinct temporal granularity and dynamics.} Streaming statistics obtained from the back-end of digital platforms exhibit a coarser granularity than multi-modal features like speech rate. Additionally, there are variations and discrepancies in granularity across these modalities, posing significant difficulties for conducting joint analysis. In summary, the comprehensive evaluation of marketing strategies in livestream e-commerce is a challenging task due to the absence of an effective approach for joint analysis of multi-modal performance and synchronous feedback, the high-dimensional and prolonged nature of livestream e-commerce data, and the distinct granularities of livestream performances and streaming statistics.

\par To address the above challenges, we have developed an interactive visual analytics system, \textit{LiveRetro}, to facilitate strategic retrospect in livestream e-commerce, and provide a platform for evidence-based research in marketing. Our work makes the following contributions. \textbf{First}, we identify seven design requirements that support a comprehensive strategic retrospect in livestream e-commerce and informative computational features that facilitate the analysis of live performance in this domain. \textbf{Second}, we present \textit{LiveRetro}, a visual analytics system designed around the identified requirements, and that incorporates enhanced visualizations and rich interaction to support the retrospective analysis of livestream e-commerce strategies from a multifaceted and empirical perspective. \textbf{Third}, we present two case studies and expert interviews that demonstrate the effectiveness of \textit{LiveRetro}. In summary, we present, \textit{LiveRetro} as the first research effort that uses a visual analytics approach to perform joint analysis of video content and synchronous feedback, aiming to decipher the intricate correlations in livestream e-commerce. Our study is intended to stimulate discussion and inspire further research into the application of visual analytics to the analysis of livestream and the livestream e-commerce domain.

\section{Related Work}
\subsection{Livestream E-Commerce Analysis}
\par Livestream e-commerce has been extensively studied across various dimensions, including its characteristics and impacts~\cite{lu2021live,sun2019live,wongkitrungrueng2020role}, customer motivations~\cite{kang2021dynamic,lee2021impulse,ming2021social,zhang2022characteristics}, and streamers' and sellers' engagement tactics~\cite{guo2022way,park2020effects,wongkitrungrueng2020live}. Some studies have focused on strategy discovery and validation~\cite{wongkitrungrueng2020live, ram2019live,wu2022malicious,wu2021forming}, such as the utilization of personalities, game-prizes, and shows as part of the Persuasion-based Approach to elicit interest and engagement from viewers~\cite{wongkitrungrueng2020live}. However, none of these studies have addressed the practical strategy execution in livestream e-commerce. This study contributes to the existing literature by identifying the influences of various performance features based on streamers' own live performance and streaming statistics, providing personalized and concrete insights.

\par The prevailing research methodology in the field typically involves empirical studies conducted through online questionnaires~\cite{guo2021effects,hu2020enhancing} and qualitative research methods~\cite{lu2020exploratory}. While some studies have attempted to address subjectivity and biases in data collection~\cite{kang2021dynamic,fei2021promoting}, they have mainly focused on monitoring limited components of livestream e-commerce, such as online counts and streamers' replies. In contrast, this study adopts a more direct approach by gathering statistics and feedback from the live streaming back-end and extracting features from live replays. This evidence-based platform enables a more comprehensive and objective analysis of the livestream e-commerce phenomenon.

\subsection{Visual Analysis of Videos}
\par In the past decade, various visual analytic approaches have been proposed to analyze different types of videos, including movies~\cite{kurzhals2016visual, ma2020emotionmap}, news~\cite{john2019visual}, presentations~\cite{zeng2019emoco}, education~\cite{shi2015vismooc,zeng2020emotioncues}, and user experience. Specifically, many researchers analyze human behavior and its dynamics in videos, including emotion~\cite{zeng2019emoco,maher2021ffective}, voice modulation~\cite{yuan2019speechlens}, body language~\cite{li2021visual,zeng2022gesturelens, wu2018multimodal}, and word use and delivery~\cite{wang2021dehumor}.
Some researchers leverage video events to evaluate user behavior, such as clickstream for learning behavior analysis in the MOOC platform~\cite{shi2015vismooc} and user comment posting~\cite{chen2022danmuvis} for studying viewers' online participation. 
Two recent studies have proposed innovative approaches for evaluating user experience. Soure et al.~\cite{soure2021coux} developed a tool called \textit{CoUX} which facilitates the review of think-aloud usability test videos. Similarly, Batch et al.~\cite{batch2023uxsense} introduced \textit{uxSense}, which utilizes machine learning techniques to extract user behavior from audio and video recordings as parallel time-stamped data streams. Notably, Tang et al.~\cite{tang2021videomoderator} proposed a risk-aware framework, \textit{VideoModerator}, for multi-modal video moderation in e-commerce.

\par However, these approaches mainly focus on exploring information within the video and are not directly applicable to analyzing livestream e-commerce data, where streamers' live performance and viewers' feedback should be considered simultaneously. Our system integrates ML modeling, explanations, and visualization to enable a joint analysis of video content and synchronous feedback.

\subsection{Time-series Forecasting and Explanatory Visualization}
\par Time-series forecasting is widely used in e-commerce to predict sales and product prices, with traditional statistical forecasting methods such as ARIMA models~\cite{zhang2003time} and 
machine learning approaches like multi-layer perception (MLP)~\cite{zhang1998forecasting} and recurrent neural networks (RNNs)~\cite{elman1990finding} being employed. Nonetheless, improved prediction performance alone is insufficient for decision-makers who require trust, adoption, and regulatory compliance. Therefore, studies have explored integrating explanatory visualizations with forecasting, such as \textit{MultiRNNExplorer}~\cite{shen2020visual} and \textit{mTSeer}~\cite{xu2021mtseer}, which enhance line charts with feature contributions to explain and evaluate multivariate time-series forecasting models. Other studies, such as 
\textit{PromotionLens}~\cite{zhang2022promotionlens} and \textit{RISeer}~\cite{chen2022riseer}, have also utilized feature contributions to explain sale volume forecasts and evaluate regional industrial structure.

\par Previous explanatory designs mainly provide feature-level explanations for predictions. To overcome the challenge of aggregating multiple feature contributions from diverse channels, we use representative ML models to model streaming statistics and generate feature contribution summaries at multiple levels. We augment the \textit{mTSeer} design by adding partitioned stacked bars to display the internal distribution and using a ``lollipop'' design to segment the summary for each merchandise based on its temporal periods.

\begin{figure}[h]
 \centering 
   \vspace{-4mm}
 \includegraphics[width=\columnwidth]{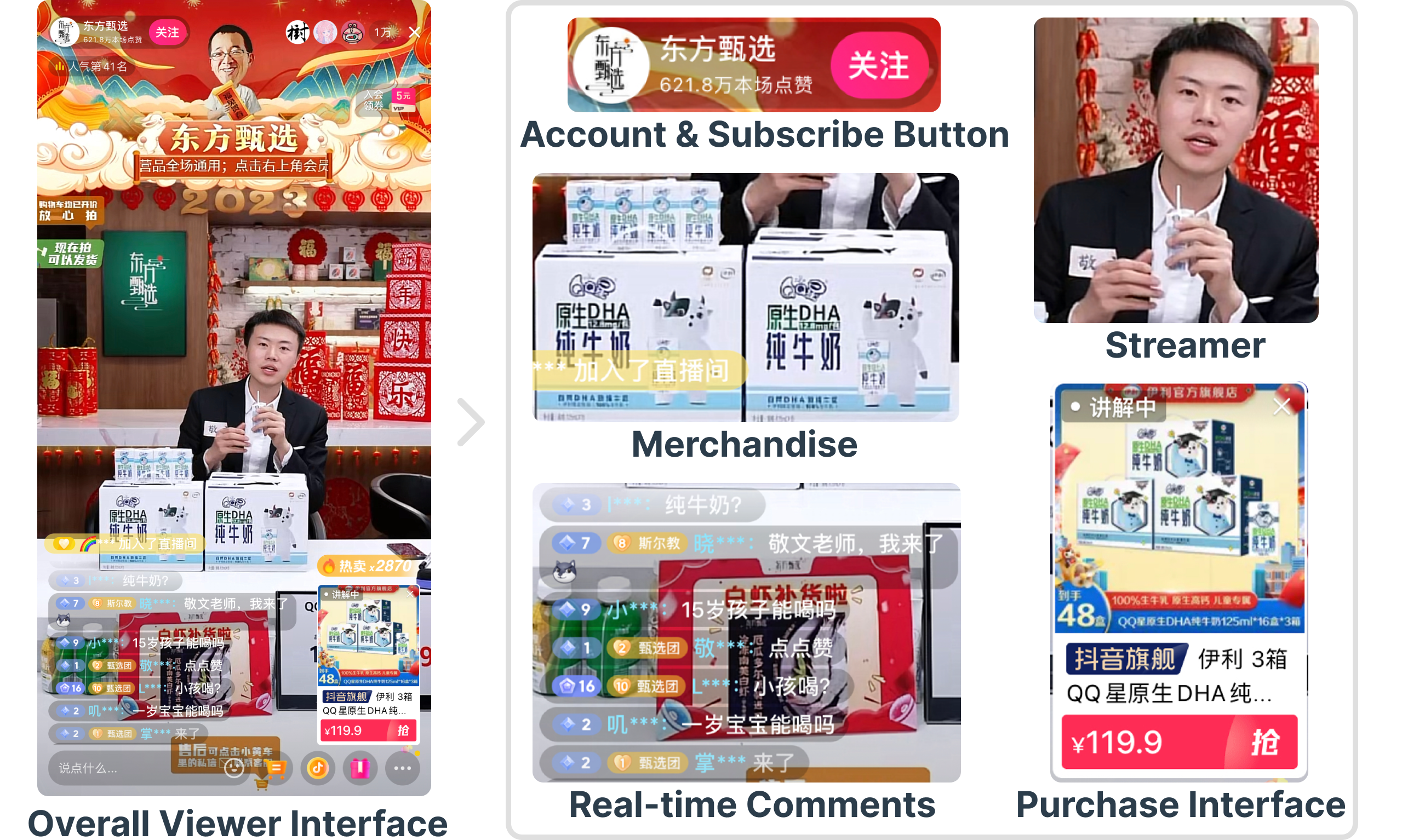}
 \vspace{-6mm}
 \caption{A viewer interface for livestream e-commerce on \textit{TikTok} with five key components: the \textit{Account}, the \textit{Streamer}, the \textit{Merchandise}, the \textit{Comments}, and the \textit{Purchase Interface}.}
 \label{fig:platform}
  \vspace{-4mm}
\end{figure}

\section{Observational Study}
\subsection{About Livestream E-Commerce Platforms}
\par Livestream e-commerce platforms, such as \textit{TikTok}, have a viewer interface (\cref{fig:platform}) that includes five key components: \textit{Account}, \textit{Streamer}, \textit{Merchandise}, \textit{Comments}, and \textit{Purchase Interface}. \textit{Account} serves as a unique identifier, \textit{Streamer} functions as salespeople, \textit{Merchandise} is the promoted product, \textit{Comments} provide real-time feedback, and the \textit{Purchase Interface} enables immediate purchases.

\subsection{About the Collaborative Team and User Interviews}
\par We worked with three experienced livestream e-commerce streamers, \textbf{\textit{E1}}, \textbf{\textit{E2}}, and \textbf{\textit{E3}}, who are among the top 1\% of sales on \textit{TikTok}. They have been using commercial analytical tools in their daily analytical routine for over 14 months. We also collaborated with two marketing researchers from a local university, \textbf{\textit{E4}} and \textbf{\textit{E5}}, who have publications in the field of marketing in livestream e-commerce. To identify appropriate analytical approaches and requirements for conducting a strategic retrospect in the context of livestream e-commerce, semi-structured independent interviews lasting approximately $40$ minutes were conducted with the five participants. The interviews aimed to elicit 1) information about the participants' current practices and bottlenecks related to strategic retrospect and marketing research, 2) specific concerns regarding livestream e-commerce strategy, and 3) proposed features for an approach that facilitates strategic retrospect in e-commerce.


\par \textbf{Current Practices.} As per the feedback from the streamers, livestream e-commerce streamers follow a specific set of steps for conducting a strategic retrospect after their live streaming session. First, they identify the highlights and mistakes of their performance during the session, such as identifying any interaction issues or mistakes that were made. Second, they comb through the streaming statistics to examine their live performance and detect any significant variations. Finally, they solicit feedback from viewers to gain insight into their reactions towards the live performance and the demand for the merchandise. In contrast, marketing researchers typically construct conceptual models and formulate hypotheses based on relevant literature, conduct surveys, and use confirmatory factor analysis~\cite{brown2015confirmatory} and structural equation modeling~\cite{ullman2012structural} to test their hypotheses and derive theoretical and practical implications based on their findings.

\par \textbf{Bottlenecks.} Commercial streaming tools offer live replay and streaming statistics for professional streamers, but users are limited to selecting specific points in time to watch the original video. \textbf{\textit{E1}} mentioned that ``\textit{I have been trying to figure out what are the most selling features, but it's pretty hard for me to just watch the videos and digest them over and over again.}'' This can make conducting strategic retrospect inefficient and challenging. Marketing researchers also face constraints when relying solely on empirical research for broader research scope and solid reasoning. ``\textit{Inductive reasoning, which forms the basis of empirical research, can obscure exploration of reality and limit the reliability and universal applicability of generalizations}'', said \textbf{\textit{E5}}. As such, streamers and researchers alike value the development of an evidence-based platform for expanding research scope, validating existing theories, and conducting comprehensive strategic retrospects.

\par \textbf{Analytical Aspects and Approaches.} Experts were interviewed individually to determine the main analytical aspects of the strategic retrospect, and a consensus was reached on three core elements: \textbf{streamer}, \textbf{viewer}, and \textbf{merchandise}. According to \textbf{\textit{E2}}, ``\textit{these are three golden elements of every live streamer}''; ``\textit{the streamer promotes certain merchandise, and the viewer is driven by the streamer's live performance to purchase the merchandise. These three form the core of the livestream e-commerce session}''. We asked the experts to provide their professional opinions on various aspects of the analysis. \textbf{\textit{E1}} highlighted the importance of live performance analysis for streaming from the streamer side. \textbf{\textit{E5}} expressed interest in exploring the impact of live performance from different channels, such as audio and frames. Therefore, we discussed and identified a multi-modal approach for live performance analysis with the researchers. Unfortunately, the literature in the livestream e-commerce field only describes the characteristics of streamers as abstract notions~\cite{zhang2022characteristics,Yang2023ImpactOS,guo2022way} (e.g., expertise, attractiveness). Therefore, we referred to relevant literature~\cite{wongkitrungrueng2020live,song2021drives} and conducted a focus group discussion with all experts to further brainstorm and identify modalities and computational characteristics of each modality. We identified several key attributes for the \textit{Audio}, \textit{Text}, and \textit{Frame} channels in the context of livestream e-commerce, including \textit{volume}, \textit{pitch}, \textit{speech rate}, \textit{pause}, six types of \textit{sales pitches}, \textit{facial expression}, and \textit{camera position} (Tab.1 in the appendix for details). It is noteworthy that the six types of sales pitches we identified are commonly utilized in livestream e-commerce, which was corroborated by all streamers, but had not been previously recognized in the literature. From the viewer side, \textbf{\textit{E3}} suggested summarizing viewer comments, while \textbf{\textit{E1}} emphasized the importance of understanding viewers' interaction status. Regarding the commodity aspect, \textbf{\textit{E2}} specified two sub-aspects, namely the selection and the permutation of commodities. In addition to these three aspects, all experts showed enthusiasm for exploring the correlation between live performance and streaming statistics. \textbf{\textit{E2}}, \textbf{\textit{E4}}, and \textbf{\textit{E5}} further suggested investigating the influences of these three aspects and more detailed features on specific targets (e.g., sales, likes).

\newtcbox{\mybox}[1][gray]
  {on line, arc = 2pt, outer arc = 3pt,
    colback = gray!10!white, colframe = white!75!black,
    boxsep = 0pt, left = 1pt, right = 1pt, top = 1pt, bottom = 1pt,
    boxrule = 1pt, bottomrule = 1pt, toprule = 1pt}

\subsection{Design Requirements}
\label{D.R.}
\par We derived seven design requirements from user interviews and expert feedback, categorized into four groups: streamer retrospect (\mybox{Streamer}), viewer retrospect (\mybox{Viewer}), merchandise retrospect (\mybox{Merchandise}), and correlation exploration (\mybox{Correlation}).
\par \noindent\mybox{Streamer} \textbf{R1. Summarize the performance of each streamer} in a live session. Livestream e-commerce often requires multiple streamers to host extended live sessions, making it essential to assess and combine each streamer's performance in terms of sales and viewer response for further analysis and reflection.
\par \noindent\mybox{Streamer} \textbf{R2. Capture features of streamers' live performance} from multiple channels. Experts agreed that relying on repetitive raw video observation to evaluate streamers' live performance is tedious and time-consuming. 
A more efficient and effective approach is required to examine streamers' live performance.
Therefore, they suggested using features from multiple channels to create a more comprehensive and detailed picture of the streamer's performance.
\par \noindent\mybox{Viewer} \textbf{R3. Summarize viewers' comments} for the chosen time period. Viewer comments during live streams offer direct feedback on the streamer and the promoted merchandise and are influenced by live room activity and streamer guidance. Therefore, it is important to synthesize and summarize viewer comments during the relevant time period to address pre-sales inquiries and post-sales concerns.
\par \noindent\mybox{Merchandise} \textbf{R4. Summarize sales data for each merchandise.} During a live session, streamers promote multiple merchandise and require access to sales data for each item to analyze and optimize their strategy. High-conversion merchandise may be promoted longer while underperforming items may be removed from the promotion.
\par \noindent\mybox{Merchandise} \textbf{R5. Visualize merchandise arrangement.} Streamers promote merchandise in a specific order during live sessions, categorizing them as ``traffic-type'' for visibility, ``welfare-type'' for interaction, and ``profit-type'' for maximum profit. An overview of the merchandise arrangement is needed to evaluate its effectiveness in driving sales.
\par \noindent\mybox{Correlation} \textbf{R6. Provide an organized joint analysis} for live replay and feedback. Experts agreed on the need to connect live performances with their corresponding feedback, which is difficult to discern from video analysis alone. Thus, the system must harmonize multi-modal features and streaming statistics and present them in an organized way.
\par \noindent\mybox{Correlation} \textbf{R7. Identify the influences of different channels and features} for a live session. \textbf{\textit{E2}}, \textbf{\textit{E4}}, and \textbf{\textit{E5}} expressed a notable interest in examining the impact of specific channels or features on streaming statistics. They aimed to determine the key elements that streamers should prioritize in order to enhance sales, such as providing detailed explanations or creating an engaging atmosphere.

\section{LiveRetro}
\par
\begin{figure}[h]
 \centering 
   \vspace{-3mm}
 \includegraphics[width=\columnwidth]{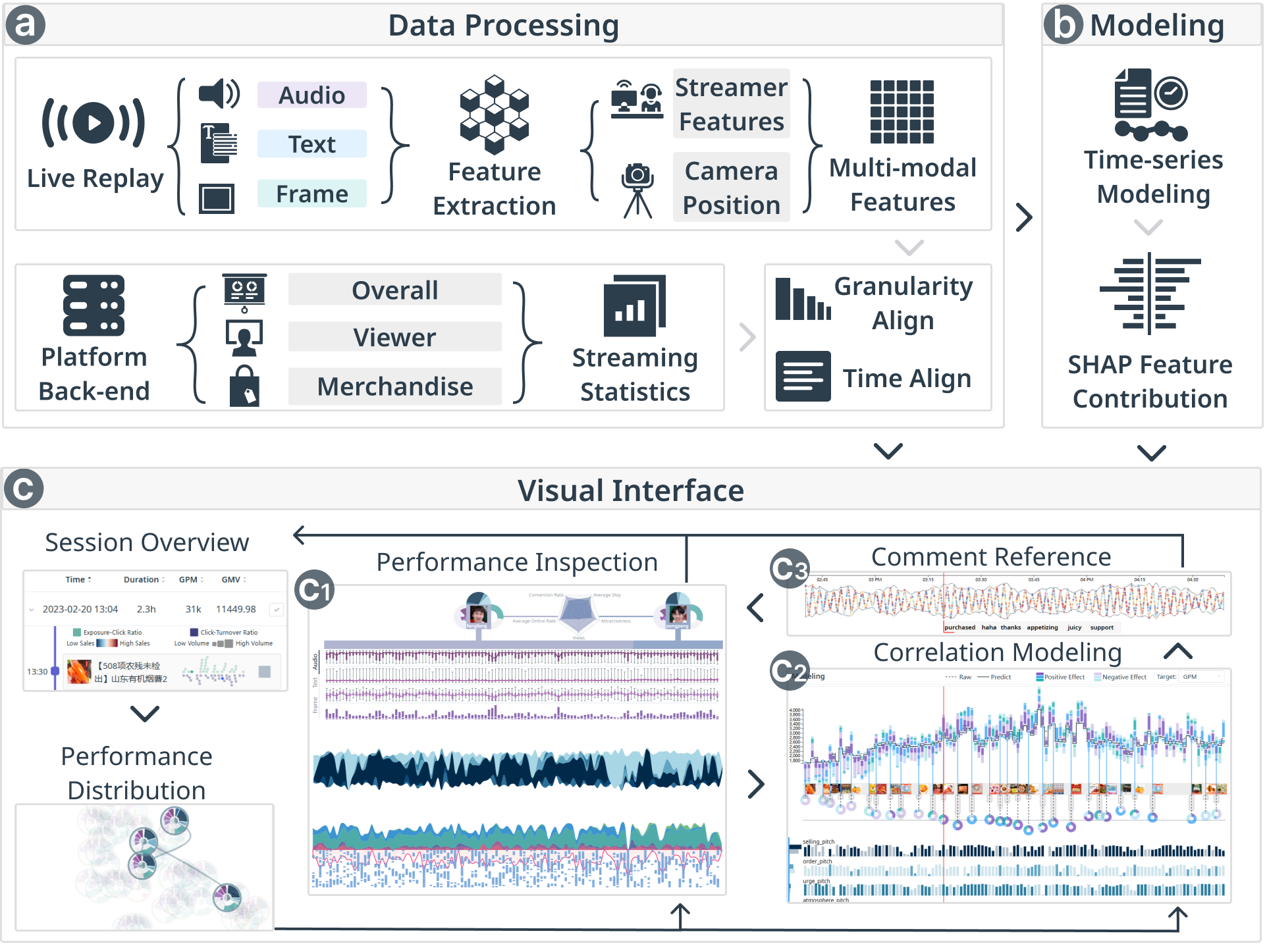}
 \vspace{-6mm}
 \caption{\textit{LiveRetro} consists of a \textit{data processing} module, a \textit{modeling} engine, and a \textit{visual interface} which supports a cyclical workflow.}
 \label{fig:approachOverview}
  \vspace{-3mm}
\end{figure}

 Our approach and workflow for analysis using the \textit{LiveRetro} system consists of three components: \darkcircled{a} data processing module, \darkcircled{b} modeling engine, and \darkcircled{c} visual interface (\cref{fig:approachOverview}). The data processing module extracts data from live replay and retrieves streaming statistics, applies models for feature extraction, and aligns features and statistics to the same granularity and time interval. The modeling engine utilizes the aligned performance features to model certain attributes with multiple models, and the model with the best performance is automatically selected. Lastly, the visual interface provides multiple interconnected views for detailed and comprehensive retrospective analysis.

\subsection{Data Description and Processing}
\par We used a dataset provided by our collaboration team members \textbf{\textit{E1}}-\textbf{\textit{E3}}, consisting of livestream e-commerce data from January 1 to February 28, 2023. This dataset comprises all live streaming sessions on their account during the time period, with corresponding videos and streaming statistics. The videos contain frames and audio capturing the sales pitch of the streamers and their display, enabling viewers to witness the promotional activity. The streaming statistics (\cref{tab:statistics}), on the other hand, provides synchronous feedback on overall, viewer, and merchandise aspects. Despite the extended duration of individual live streaming sessions, which typically surpass ten hours, streamers endorse a range of approximately 30 distinct products through their shift work. In consultation with \textbf{\textit{E1}} and \textbf{\textit{E2}}, we segmented each session into multiple clips, each featuring a complete batch of products.

\par In this study, we examined three modalities of video data, which include audio, text, and frame. To extract features from these modalities, we adopted a two-step process. First, we separated the audio from the video and then utilized the Google Cloud Speech-to-Text API to transcribe the audio data into script form. The resulting script was then aligned with the time series data to facilitate feature extraction. To construct our training dataset, we selected script data from five different livestream sessions that correspond to the video data. Prior to the following feature classification task, we preprocessed the data by removing incomplete sentences as they can adversely affect classification accuracy. Additionally, we corrected any inaccuracies resulting from the transcription process. Overall, our approach involved the extraction of features from the various modalities of video data, followed by the alignment of these features with the corresponding time series data, enabling us to accurately classify and analyze the features of interest.

\begin{table}[h]
\vspace{-3mm}
\centering
\caption{Time-Series Feedback Retrieved from Platform Back-End.}
\label{tab:statistics}
\vspace{-3mm}
\resizebox{\columnwidth}{!}{%
\begin{tabular}{ccccccc}
\hline
\multicolumn{7}{c}{\textbf{Overall}}                                                                                               \\ \hline
\multicolumn{2}{c|}{Sales}   & \multicolumn{2}{c|}{Volume}               & \multicolumn{1}{c|}{GPM} & \multicolumn{2}{c}{UV Value} \\ \hline
\multicolumn{7}{c}{\textbf{Viewer}}                                                                                                \\ \hline
\multicolumn{1}{c|}{Entry} &
  \multicolumn{1}{c|}{Departure} &
  \multicolumn{1}{c|}{Like} &
  \multicolumn{1}{c|}{Comments} &
  \multicolumn{1}{c|}{Subscribe} &
  \multicolumn{1}{c|}{Conversion Rate} &
  Average Stay \\ \hline
\multicolumn{7}{c}{\textbf{Merchandise}}                                                                                           \\ \hline
\multicolumn{2}{c|}{Cancels} & \multicolumn{2}{c|}{Exposure-Click Ratio} & \multicolumn{3}{c}{Click-Turnover Ratio}                 \\ \hline
\end{tabular}%
}
\vspace{-3mm}
\end{table}

\subsection{Feature Extraction}
\label{sec:featureExtraction}
\par By conducting interviews with streamers (\textbf{\textit{E1}}-\textbf{\textit{E3}}) and researchers (\textbf{\textit{E4}}–\textbf{\textit{E5}}), we identified the requisite features pertaining to the following three modalities. These features facilitate a more comprehensive comprehension and analysis of the streamer's information, thereby allowing us to gain a better understanding of the strategies employed.

\par \textbf{Audio Feature Extraction.} Our analysis of the audio data centers around four prominent features, namely \textbf{Volume}, \textbf{Pitch}, \textbf{Pause}, and \textbf{Speech Rate}, as described in Tab.1 in the appendix. We selected these features due to their capacity to encapsulate a diverse range of audio characteristics. We employed \textit{Praat}~\cite{boersma2001praat} to directly obtain Volume and Pitch from its audio analysis view. To extract Speech Rate, we utilized a \textit{Praat} script that identified the occurrence time and the number of syllables. We then computed the syllable count per second to derive the Speech Rate at one-second intervals.

\par \textbf{Text Feature Extraction.} Following consultation with both streamers and researchers, our collaboration team has reached a consensus to categorize the sales pitches employed during live streaming into six categories. These include the \textbf{Traffic Pitch}, \textbf{Interaction Pitch}, \textbf{Selling Pitch}, \textbf{Order Pitch}, \textbf{Urge Pitch}, and \textbf{Atmosphere Pitch}, as shown in Tab.1 in the appendix. With the assistance of \textbf{\textit{E1}}-\textbf{\textit{E4}}, we were able to correctly identify and categorize sentences into their corresponding sales pitch categories. The training dataset was composed of $3,000$ labeled sentences, with $500$ sentences allocated to each category. We examined two powerful pre-trained language models, BERT~\cite{devlin2018bert} and ERNIE3.0~\cite{sun2021ernie}, known for their exceptional performance in Chinese language tasks, to classify live-stream sales pitches. Both models were trained and fine-tuned using labeled sentences, and a 5-fold cross-validation technique was employed to identify the optimal hyperparameters. Given the relatively small size of our dataset and the concern of overfitting, we took careful measures to ensure that the models performed consistently on both the training and test sets. Upon validation, ERNIE3.0 achieved a higher accuracy of $91.2\%$ on the test data, outperforming BERT, which achieved an accuracy of $85.4\%$. As a result, ERNIE3.0 was selected and successfully implemented for the classification of sentences extracted from five live-stream sessions.

\par \textbf{Frame Feature Extraction.} The selection of \textbf{Facial Expressions} and \textbf{Camera Positions} as streamer video frame features is a thoughtful process. Facial expressions can provide insights into the streamer's personality and significantly impact viewer perception. For example, a genuine smile can create a warm and friendly impression, enhancing the streamer's persuasiveness and credibility. Additionally, camera positions can showcase a product from different angles and locations, enhancing audience understanding and maintaining their engagement. In frame segmentation of livestream videos, the \textit{RetinaFace} algorithm~\cite{deng2020retinaface} is employed for accurate face detection, given its excellent performance in this task. After face detection, the \textit{VGG-Face} model~\cite{parkhi2015deep} is utilized to extract emotional expressions from frames containing human faces. Considering the similarities between camera positions in film and the unique characteristics of livestream e-commerce, frames are classified into two categories: \textit{Long-Range View} and \textit{Close-Up}, as detailed in Tab. 1 in the appendix. The classification is based on the results of face detection and the proportion of the face area in each frame.

\subsection{Time-series Modeling}
\label{sec:modeling}
\par The extracted features were utilized as input for time-series modeling, with the features aggregated at a one-minute granularity to facilitate analysis. The feature vector consisted of $21$ dimensions, including six multi-modal features and $15$ real-time feedback features such as sales amount, sales volume, number of comments, and viewer average stay. To enable users to select the target data to be predicted, nine options were provided, including \textit{sales amount}, \textit{sales volume}, \textit{UV value} (sales amount divided by the number of unique viewers), \textit{GPM} (transaction amount per thousand views), \textit{number of viewers entering and leaving}, \textit{number of likes}, \textit{number of comments}, and \textit{average stay time}. Users could choose one option for prediction while using the rest as input data for prediction. To assess the model's performance, absolute and relative metrics, namely \textit{mean absolute error (MAE)} and \textit{mean absolute percentage error (MAPE)}, were utilized due to the varying values of the feature vector. Several non-linear machine learning models, including \textit{XGBoost}~\cite{chen2016xgboost}, \textit{Random Forest (RF)}, \textit{Long Short-Term Memory Networks (LSTM)}~\cite{graves2012long}, and \textit{Multi-Layer Perceptron (MLP)}, were evaluated using a 7:3 split of training and test data. The MAE and MAPE were calculated on the test data and combined to obtain a weighted average. The model with the best performance was selected based on the composite score, considering both prediction accuracy and interpretability. The chosen model was then utilized to generate predictions for the specific target data. The objective of the modeling is to examine the impacts of various channels and features. For this purpose, \textit{SHapley Additive exPlanations (SHAP)} values~\cite{lundberg2017unified} were employed as the model explanation metrics due to their suitability, consistency, and flexibility. This approach allows users to explore feature contributions at different levels and make informed decisions, such as optimizing their strategies based on identified feature pairs in streamers and merchandise.

\section{Visual Interface}
\newtcbox{\dpbox}[1][gray]
  {on line, arc = 1pt, outer arc = 2pt,
    colback = blue!50!green, colframe = blue!50!green,
    boxsep = 0pt, left = 1pt, right = 1pt, top = 1pt, bottom = 1pt,
    boxrule = 1pt, bottomrule = 1pt, toprule = 1pt}

\par Considering the large and complex data generated by livestream e-commerce, we probed into the behavioral characteristics of retrospective analysis with our collaborative team and developed a refined set of design principles, in addition to following the visualization mantra of ``overview first, zoom and filter, then details on demand''~\cite{shneiderman1996eyes}. These principles were integrated into the design process to simplify the system's visual complexity and reduce users' cognitive load during the joint analysis of extended videos and concurrent feedback.

\par \dpbox{\textcolor{white}{\small{\texttt{DP1}}}} \textbf{Consistent encoding pattern.} 
Given the complexity of livestream e-commerce data, users may experience a considerable cognitive burden in processing it. Therefore, to alleviate this burden, a consistent attribute encoding pattern can be employed across the system. This pattern entails using the same or similar shapes, positions, and color groups (e.g., \audio{purple for audio}, \text{blue for text}, and \framekk{green for frame}) at the \textbf{attribute level} for elements sharing the same attribute.

\par \dpbox{\textcolor{white}{\small{\texttt{DP2}}}} \textbf{Global visual alignment.} Users may need to refer to multiple metrics or statistics simultaneously to conduct a comprehensive analysis of livestream e-commerce strategies. This process can be simplified by globally aligning the presentation of different elements at the \textbf{inter-view level}, enabling users to compare attributes in parallel. For instance, in our \textit{Exploration View}, we align the presentation of various elements with the timeline, which features a vertically synchronized cursor to help users locate different attributes simultaneously.

\par \dpbox{\textcolor{white}{\small{\texttt{DP3}}}} \textbf{Cross-view time linkage.} We create multiple views with different objectives and capabilities to meet our design goals, providing users with a comprehensive perspective during strategic retrospect. These views may need to be navigated to access various information for a given livestream session. At the \textbf{intra-view level}, time segments are used to connect all the views, with each view focusing on elements corresponding to the desired time segment, as described in \cref{sec:interaction}.

\subsection{Session View}
\label{sec:sessionView}
\par The \textit{Session View} (\cref{fig:teaser}-\circled{A}) provides an overview of live sessions within a specific time range, along with sales data for the merchandise promoted during each session (\textbf{R5}). This view presents a table consisting of four columns that detail fundamental information and key indicators, namely the \textit{time}, \textit{duration}, \textit{GPM}, and \textit{GMV (Gross Merchandise Volume)}. The display of these four attributes allows users to obtain a swift summary of each live session. Additionally, users have the option to sort sessions according to these attributes by clicking \includegraphics[height=1.5ex]{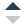} on the header. Users can expand the row of a specific session to view details of the merchandise by clicking \includegraphics[height=1.5ex]{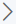} on the left. Sales data for each promoted merchandise is presented in chronological order within an expanded table along a vertical timeline, with the launch time displayed on the left and sales details grouped in boxes on the right. Each box lists the thumbnail and title of the merchandise on the left, while on the right, the overall Exposure-Click Ratio (i.e., the click number of the purchase interface to online counts) and Click-Turnover Ratio (i.e., the number of transactions to click count of the purchase interface) distribution is depicted using two opposing beeswarm charts. Highlighted dots indicate the exact value of two attributes for the merchandise. The size of the rectangle on the right encodes the volume, while the color encodes sales. In terms of design, the amount of sales for each merchandise follows the same encoding scheme as the \textit{Exploration View} (\dpbox{\textcolor{white}{\small{\texttt{DP1}}}}). Lastly, to facilitate comparison, the area graphs and sales \& volume rectangles are aligned vertically (\dpbox{\textcolor{white}{\small{\texttt{DP2}}}}).

\subsection{Segment View}
\label{sec:segmentView}
\par The \textit{Segment View} (\cref{fig:teaser}-\circled{C}) allows users to investigate the temporal distribution of streamers' live performance (\textbf{R2}) for different time segments. Additionally, it provides a visual representation of the relationship between \textit{GPM} and live performance (\textbf{R7}), conveying this information in an intuitive manner.
\par Inspired by \textit{time curves}~\cite{bach2015time}, to unfold the temporal variation pattern in live performance, we partition live sessions into time segments of varying granularities (i.e., 1-minute and 5-minute) and project performance features (detailed in Tab.1 in the appendix) of each segment as a glyph onto a 2D plane using \textit{t-SNE}, because it offers ``\textit{the best overall quality in terms of producing low errors on average}''~\cite{espadoto2019toward}. To construct the feature vector, we extract each audio feature by taking its minimum, median, and maximum values (e.g., $V_{volume}=[Volume_{min},Volume_{median},Volume_{max}]$), text features are extracted by calculating the total number of words in each pitch's sentence, and facial features are extracted by determining the primary expression category and its frequency. We normalize each group of features before concatenation them to prevent bias, resulting in a 25-dimensional vector. Each selected glyph point is then connected by curves in chronological order, with the hue of the curves varying.

\par To effectively display the performance features of each time segment, we design a glyph based on a pie chart. Within the glyph, the central pie chart displays the relative value of \textit{GPM}. Both the pie's and outer ring's colors represent the chronological order. To encode audio, text, and facial features, this glyph is equally divided into three sectors of a circle. Specifically, the top-left sector represents audio features, the top-right sector represents text features, and the bottom sector represents facial features. As these features are heterogeneous, we adopt different designs to differentiate each group of features. Audio features are encoded in a rose chart-based design, where the petal length represents the feature's average value of the time segment. Text features are encoded in a pie chart-based design, where the angle of the pie represents the relative amount of corresponding pitch in the time segment. Finally, facial features are encoded in a donut chart-based design, where the color represents the type of primary expression and the angle of the arc represents the intensity.

\begin{figure}[h]
 \centering 
 \vspace{-3mm}
 \includegraphics[width=0.85\columnwidth]{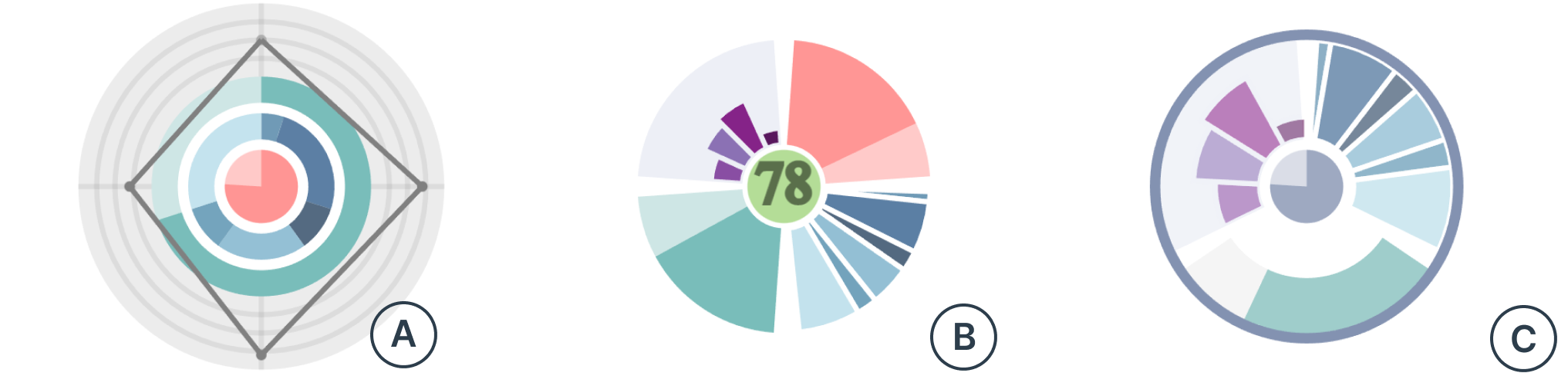}
 \vspace{-3mm}
 \caption{Glyph designs for \textit{Segment View}. (A) An alternative based on radar chart and donut chart; (B) An alternative based on pie chart and rose chart; (C) Our current design.}
 \label{fig:glyph_alternative}
  \vspace{-3mm}
\end{figure}

\par \textbf{Design Alternatives.} During the iterative design process, we explored two alternative designs (\cref{fig:glyph_alternative}-\circled{A}\circled{B}) for our glyph-based visualization. The first alternative design employed a combination of a radar chart and a donut chart, with the central pie chart representing the \textit{GPM}, and two concentric donut charts representing the proportions of sales pitches and facial expressions. Additionally, the outer radar chart was used to encode the value of four audio features. However, this alternative design lacked a clear representation of the time attribute and suffered from visual obscurity. In the second alternative design, we used a pie chart and a rose chart to represent the chronological order and the \textit{GPM} and features from three channels, respectively. However, our users reported that it was difficult to distinguish between the feature groups. This led us to explore different design options for the features from the three channels, resulting in the current design (\cref{fig:glyph_alternative}-\circled{C}). In this design, we use different visual encodings to represent the audio, visual, and textual features, respectively.

\subsection{Exploration View}
\label{sec:explorationView}
\par The \textit{Exploration View} contains \textit{Performance Inspection}, \textit{Correlation Modeling}, and \textit{Comment Reference} (\textbf{R1}-\textbf{R4}, \textbf{R6}-\textbf{R8}), aligned vertically with a synchronized cursor (\dpbox{\textcolor{white}{\small{\texttt{DP2}}}}) for concurrent analysis.

\subsubsection{Performance Inspection}
\par The \textit{Performance Inspection} part (\cref{fig:approachOverview}-\darkcircled{\small{C}\raisebox{0.4ex}{\textsubscript{1}}}) facilitates users to inspect the overall performance of streamers (\textbf{R1}) as well as their fine-grained performance across multiple channels, including audio, text, and frame (\textbf{R2}). Both representations of overall performance and fine-grained performance are also vertically aligned (\dpbox{\textcolor{white}{\small{\texttt{DP2}}}}).

\par At the top is a legend panel that dynamically switches its legends according to the selected channel. Adjacent to the legend panel is a button group. Below the legend panel, a tree metaphor is employed to summarize the performance of streamers. The ``root'' denotes the time a streamer has occupied, while the ``trunk'' displays their portraits and names. The three ``branches'' represent summaries of streamers' audio features, text features, and facial features, respectively, in a clockwise arrangement. Importantly, these summaries are encoded using detached glyph patterns that are aligned in similar positions to those in the \textit{Segment View} (\dpbox{\textcolor{white}{\small{\texttt{DP1}}}}). If multiple streamers host the session, a radar chart is displayed to compare their performance across multiple metrics, including \textit{Average Online Rate}, \textit{Views}, \textit{Attractiveness}, \textit{Average Stay}, and \textit{Conversion Rate}. Finally, at the bottom of the display, the fine-grained performance of the aforementioned channels is presented.
\par \textbf{\textit{Audio.}} Inspired by \textit{SpeechLens}~\cite{yuan2019speechlens}, we explored the use of volume charts to encode audio features. However, due to the rapidly changing dynamics of audio features, such as volume and pitch, a coarse-grained representation proved insufficient in capturing the nuances of variations within the time segments. Consequently, we opted to use connected box charts instead (\cref{fig:teaser}-\darkcircled{1}), which allowed us to represent the variations in volume, pitch, and speed in a more detailed and comprehensive manner. Additionally, a bar chart was included at the bottom of the visualization to depict the total number of pauses in each time segment, with a white line indicating the maximum pause duration.
\par \textbf{\textit{Text.}} The streamgraph chart (\cref{fig:teaser}-\darkcircled{2}), a prevalent technique for visualizing continuous data, has been employed in our study to illustrate the temporal distribution of sales pitches. Its fluid and cohesive shape, along with its suitability for analyzing the relative proportions of the entire dataset, make it a particularly fitting choice for this purpose.
\par \textbf{\textit{Frame.}} Given that the camera positions are integral to the captured facial expressions, we have presented two opposing visualization side-by-side. The stacked area chart positioned above portrays the temporal evolution and variability in the quantity and dominant type of sales pitches, while the adjacent area chart below depicts the total number of instances in which the camera was focused in a \textit{Close-up} view, with the corresponding timestamps arranged in columns from top to bottom.
\par Furthermore, users have the ability to navigate between these channels by selecting the respective tabs located on the left-hand side.

\subsubsection{Correlation Modeling}
\par
\begin{figure}[h]
 \centering 
 \vspace{-3mm}
 \includegraphics[width=\columnwidth]{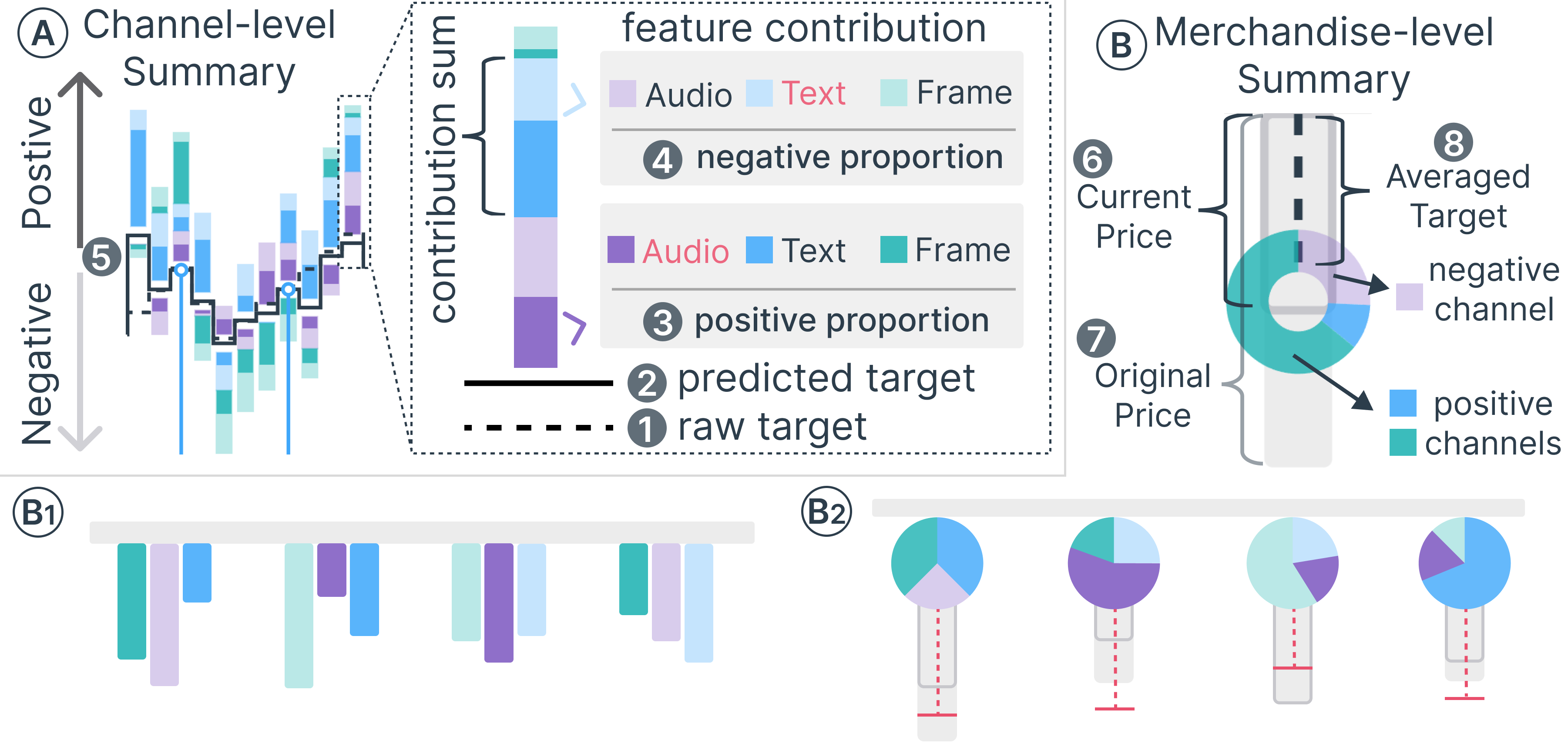}
 \vspace{-6mm}
 \caption{Contribution summary designs for \textit{Correlation Modeling} part. (A) Design for the channel-level summary; (B) Design for the merchandise-level summary. (B1) An alternative of merchandise-level summary based on bar groups design. (B2) An alternative of merchandise-level summary based on a ``cylinder''-like design.}
 \label{fig:summary_detail}
  \vspace{-3mm}
\end{figure}

 The \textit{Correlation Modeling} part (\cref{fig:approachOverview}-\darkcircled{\small{C}\raisebox{0.4ex}{\textsubscript{2}}}) provides multi-level summaries of feature contribution, enabling a thorough understanding of channel and feature effects (\textbf{R8}). It is organized in a top-to-bottom order. First, users have the option to select a specific target for inspection by utilizing the selection box located at the top right corner of the interface. In the channel-level summary (\cref{fig:summary_detail}-\circled{A}), we have presented the feature contribution of audio features, text features, and frame features (Fig.1 in the appendix), as the contribution of the corresponding channels. Regarding the design, the dotted step line (\cref{fig:summary_detail}-\darkcircled{1}) represents the raw value of the selected target, while the solid step line (\cref{fig:summary_detail}-\darkcircled{2}) represents the predicted value of the target in comparison to the raw value. Inspired by \textit{mTSeer}~\cite{xu2021mtseer} and \textit{PromotionLens}~\cite{zhang2022promotionlens}, we normalized and visualized the contribution of the four aspects using stacked bars. Similar to their approach, we positioned the stacked bars near the step line, where bars above the line represent a positive effect on the target, and vice versa (\cref{fig:summary_detail}-\darkcircled{5}). Each stacked bar consists of a solid and translucent part, representing the positive and negative proportions (\cref{fig:summary_detail}-\darkcircled{3}\darkcircled{4}) of feedback in the corresponding channel, respectively. Users can interact with the legends in the header to focus on specific channels, and they can click on bars for detailed insights on channel contributions at different levels. Below the channel-level summary is a chronological arrangement of products based on their respective launch times. Each merchandise thumbnail is linked to its corresponding point on the predicted step line, thereby enabling a more streamlined identification of the temporal duration occupied by each merchandise. The part beneath the merchandise arrangement exhibits merchandise-level summary (\cref{fig:summary_detail}-\circled{B}). The bar below each product thumbnail signifies the price (\cref{fig:summary_detail}-\darkcircled{6}\darkcircled{7}), while the donut chart provides a condensed summary of the contribution of the product's occupied time. The transparency and angle of the arc encode the polarity (positive or negative) and relative proportion of the summarized effect of a particular channel. Additionally, a dotted line (\cref{fig:summary_detail}-\darkcircled{8}) connects the thumbnail and donut chart, with its length indicating the averaged target value of the occupied time. Furthermore, the encoding color, transparency, and line style used to represent merchandise-level summary is identical to that used in the overview-level (\dpbox{\textcolor{white}{\small{\texttt{DP1}}}}), and these three parts are also vertically aligned (\dpbox{\textcolor{white}{\small{\texttt{DP2}}}}). The bottom part shows more detailed contribution summaries for specified channels at the feature-level and segment-level. The left section presents clusters of bars exhibiting the feature-level summary (\cref{fig:teaser}-\darkcircled{5}), which is derived by aggregating the positive and negative contribution values of the relevant indicator separately, for all time segments. On the right, each row of bars indicating the segment-level summary (\cref{fig:teaser}-\darkcircled{6}) for the corresponding feature. It is noteworthy that the color of the leftmost line denotes the selected channel, and the color scheme for feature-level and segment-level summaries are identical to that of the \textit{Performance Inspection} part (\dpbox{\textcolor{white}{\small{\texttt{DP1}}}}).

\par \textbf{Design Alternatives.} We explored two alternative designs for our merchandise-level summary. The first alternative (\cref{fig:summary_detail}-\circled{\small{B}\raisebox{0.4ex}{\textsubscript{1}}}) displayed the contribution of three channels for each merchandise as bar chart groups. However, this design did not effectively show the proportional relationship between each channel and lacked price information and averaged target details. Hence, we pursued a second alternative (\cref{fig:summary_detail}-\circled{\small{B}\raisebox{0.4ex}{\textsubscript{2}}}) using a "cylinder"-like design, with the upper pie indicating contribution proportion and the bar below representing the price. The positions of the red "pistons" displayed the averaged target of each merchandise. However, user feedback revealed difficulties in capturing the variation of averaged targets using this representation. Consequently, we established a connection between the contribution representation and the averaged target and chose to use a donut chart instead of a pie chart to minimize visual obstruction. These modifications resulted in our current design, which aims to provide a comprehensive and informative summary of the merchandise-level data.

\subsubsection{Comment Reference}
\par
\begin{figure}[h]
 \centering 
   \vspace{-3mm}
 \includegraphics[width=\columnwidth]{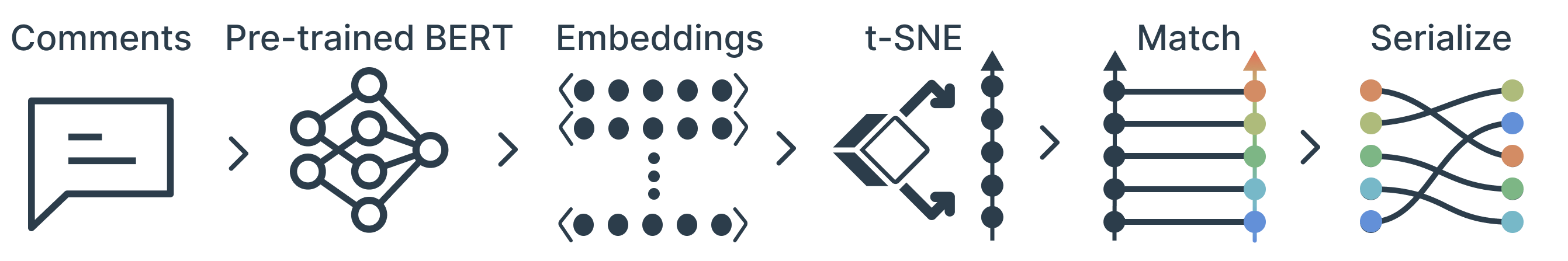}
 \vspace{-6mm}
 \caption{Comment dots are colored by creating embeddings using pre-trained BERT, projecting them onto a 1D axis with t-SNE, and assigning colors from a gradient palette.}
 \label{fig:comment_coloring}
  \vspace{-3mm}
\end{figure}

 The \textit{Comment Reference} part (\cref{fig:approachOverview}-\darkcircled{\small{C}\raisebox{0.4ex}{\textsubscript{3}}}) facilitates users in retrospectively accessing the real-time comments that were generated during live streaming, as articulated in requirements \textbf{R3} and \textbf{R6}. It includes a timeline and a summary of comments. In order to present both the overarching temporal patterns of the comments and informative details contained therein (such as opinion on merchandise), the summary is presented using a volume chart-based design with four components: a background volume chart, a zig-zag curve, a foreground with individual comment data points, and a floating keyword box. The background volume chart illustrates the trend of comment intensity, and a zig-zag curve that is bounded by the volume chart is constructed along the timeline. To elucidate the semantic distribution of the comments (\cref{fig:comment_coloring}), we first create embeddings for all comments within the clip using a pre-trained BERT model. Subsequently, we project these embeddings onto a 1D virtual axis using \textit{t-SNE}, and assign colors to each dot using a gradient palette, from left to right along the virtual axis. Finally, all colored dots are mapped back onto the zig-zag curve in chronological order. Furthermore, as the vertical cursor hovers over a particular time segment, the keyword box at the bottom floats synchronously, exhibiting the extracted keywords within that time segment in descending order of weight from left to right. Notably, the detailed matrix and comment summary are also aligned vertically with the timeline (\dpbox{\textcolor{white}{\small{\texttt{DP2}}}}).

\par
\begin{figure}[h]
 \centering 
 \vspace{-3mm}
 \includegraphics[width=\columnwidth]{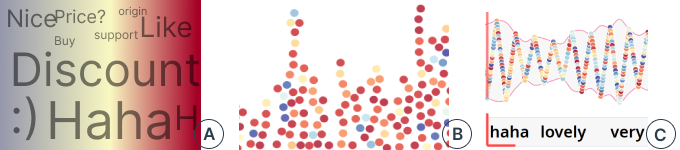}
 \vspace{-6mm}
 \caption{Comment summary designs for \textit{Comment Reference} part. (A) An alternative based on word cloud and colorfield design; (B) An alternative based on beeswarm chart; (C) Our current design.}
 \label{fig:comment_alternative}
  \vspace{-6mm}
\end{figure}

 \textbf{Design Alternatives.} We explored two alternatives for comment summary design. The first alternative (\cref{fig:comment_alternative}-\circled{A}) utilized a word cloud with a \textit{colorfield} background to indicate comment intensity along the timeline. However, this design did not effectively convey accurate semantic information for certain time segments. The second alternative (\cref{fig:comment_alternative}-\circled{B}) employed a beeswarm chart with colored dots representing the semantic distribution. While this design captured semantic variations within and between time segments, it did not reveal the sequence of comments. Consequently, the current design (\cref{fig:comment_alternative}-\circled{C}) was developed by combining the two alternatives. In this design, the dots representing comments are arranged along a zig-zag curve to maintain chronological order, and a floating keyword box is positioned below the curve to facilitate the identification of the most relevant keywords.

\subsection{Interaction Workflow}
\label{sec:interaction}
\par Incorporating a \textit{Time Linkage Mechanism}, \textit{LiveRetro} synchronizes multiple views within specific time segments (\dpbox{\textcolor{white}{\small{\texttt{DP3}}}}). This mechanism seamlessly integrates interactive operations among the views in the cyclical workflow. \textit{LiveRetro} offers users multiple options to activate the time linkage features, including \textbf{clicking} on a merchandise box in the \textit{Session View}, \textbf{lassoing} or clicking glyph points in the \textit{Segment View}, and clicking or \textbf{brushing} in the \textit{Exploration View}. Upon selecting time segments, the system responds by locating and highlighting the corresponding merchandise in the \textit{Session View}, highlighting the corresponding glyph points and time curves in the \textit{Segment View}, moving the vertical cursor in the \textit{Exploration View} to the corresponding segment, and causing the live replay to jump to the corresponding time. As shown in \cref{fig:approachOverview}-\darkcircled{c}, to complete the workflow cycle, users first select a specific time range in the \textit{Session View}, then proceed to examine merchandise sales data and designate a particular clip by clicking  \raisebox{-0.4ex}{\includegraphics[height=2.0ex]{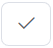}}. Next, users transition to the \textit{Segment View} to determine segment granularity and analyze performance distribution through lassoing and \textbf{semantic zooming} techniques. Subsequently, users navigate to the \textit{Exploration View} to further inspect performance using \textbf{hovering and tooltips}. Following the modeling and reference parts, users can \raisebox{-0.5ex}{\includegraphics[height=2.2ex]{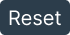}} the current selection or save selected segments as \raisebox{-0.6ex}{\includegraphics[height=2.2ex]{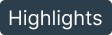}} or \raisebox{-0.5ex}{\includegraphics[height=2.2ex]{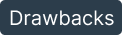}} by clicking the top right buttons. Finally, users can brush or click the \textit{Exploration View} to proceed to the next circulation back to the \textit{Session View} or to move to the \textit{Record View} for a summary of the entire process.

\par \textbf{Record View }(\cref{fig:teaser}-\circled{D}). To meet user feedback requesting follow-up references, we introduced a summarized performance feature with four attributes: corresponding target, selected time segments, summarized performance using the \textit{Segment View} and \textit{Exploration View} glyph pattern (\dpbox{\textcolor{white}{\small{\texttt{DP1}}}}), and vertically aligned timeline thumbnails for easy comparison (\dpbox{\textcolor{white}{\small{\texttt{DP2}}}}). Users can delete specific selections with the \raisebox{-0.4ex}{\includegraphics[height=2.0ex]{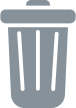}} icon and export all preserved selections, along with performance information, for further analysis \raisebox{-0.6ex}{\includegraphics[height=2.2ex]{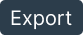}}.


\section{Evaluation}

\subsection{Case I: Strategic Retrospect for Streamers}
\par In Case I, a retrospective analysis was conducted on the livestream sessions delivered by streamer \textbf{\textit{E1}}. The focus of the analysis was centered around the streamer's efforts to reflect on past failures and employ appropriate promotional techniques for maximum efficacy.

\par \textbf{``What were the reasons for the failure of the entertainment?''} Following the interaction workflow, \textbf{\textit{E1}} opted to investigate the livestream sessions from the past ten days. \textbf{\textit{E1}} selected the session with the highest \textit{GPM} value (\cref{fig:teaser}-\casebox{1}) and examined the merchandise sales data in the \textit{Session View}. It was discovered that the merchandise ``Marker Pen'' had poor sales performance \casebox{2}, with low \textit{Exposure-click ratio} and \textit{Click-turnover ratio}. Curious about the reason for this, \textbf{\textit{E1}} accessed the corresponding period's merchandise box to view the \textit{Segment View} with the one-minute granularity selected, it was observed that the \textit{atmosphere pitch} was dominant in \textit{Text} channel during that period \casebox{3}. This prompted his interest in exploring the relationship between these features and sales data. To gain a better understanding of the sales data, \textbf{\textit{E1}} turned to the \textit{Exploration View} to investigate further. It was realized that the \textit{sales amount} and \textit{sales volume} in the \textit{Session View} were not entirely comprehensive due to the varying prices of different products. Thus, \textbf{\textit{E1}} opted to utilize \textit{GPM} as the target in the \textit{Correlation Modeling} part. However, during that period, a low \textit{GPM} value \casebox{4} was still obtained with significant negative contributions from the \textit{Text} and \textit{Frame} channels displayed in the corresponding merchandise-level summary \casebox{5}. To understand this further, \textbf{\textit{E1}} turned to the \textit{Performance Inspection} part to examine these multi-modality features. It was observed that there were very few facial expressions during this period, and a large number of \textit{close-ups} in the \textit{camera position} \casebox{6}. After analyzing the video content, it was discovered that the streamer was drawing a small rabbit on paper to convey New Year’s greetings to the viewers \casebox{7}, with many \textit{atmosphere pitches} in the \textit{Text} features. It was believed that this ``talent show'' captured the viewers' attention, leading to fewer orders.

\par Drawing upon this assumption, \textbf{\textit{E1}} formulated a hypothesis that the feedback provided by the viewer would be favorable. As a result, \textbf{\textit{E1}} modified the target data to reflect a positive sentiment by replacing it with the term \textit{Like}, and subsequently observed a peak in the \textit{Like} value during that time period \casebox{8}. Additionally, the inclusion of \textit{Text} features appeared to have a positive impact on this outcome. \textbf{\textit{E1}} conducted further examination of the \textit{Correlation Modeling} part by clicking on the blue stack bar to direct to the detailed contribution summaries below, which led to the discovery that the \textit{atmosphere pitch} has a positive contribution to the target data \casebox{9}, as determined by its \textit{SHAP} values. In order to gain a more thorough understanding,  \textbf{\textit{E1}} also investigated the situation when the target was \textit{Departure} \casebox{10}. This yielded a comparable result, with the exception of a dip in the departure value and negative contribution of the \textit{atmosphere pitch} \casebox{11}, which was sensible given that low \textit{departure} values indicate high popularity. Finally, \textbf{\textit{E1}} analyzed the \textit{Comment Reference} part \casebox{12} and found that the abundance of semantically similar comments, such as those containing terms like ``\textit{support, talented, excellent, like}'', during this time period. The prevalence of such comments suggests that viewers responded positively to the content. This further reinforced \textbf{\textit{E1}}'s initial belief. Upon analyzing the data of the three targets, it was determined by \textbf{\textit{E1}} that the use \textit{atmosphere pitch} should be moderate, as excessive employment may result in increased viewership but decreased sales. Consequently, a decision was made to maintain a balance in the utilization of various sales pitches during subsequent sessions. For ease of future reference, \textbf{\textit{E1}} appended this period to the \textit{Record View} under the category of \textit{Drawbacks} \casebox{13}.

\par \textbf{``Are there alternative methods that would yield better outcomes?''} After examining the \textit{Drawbacks}, \textbf{\textit{E1}} developed an interest in the methods that led to the high sales periods. As such, he turned his attention to the \textit{GPM} and scrutinized the raw value peaks, uncovering a surprising period where the \textit{GPM} value initially decreased but quickly rebounded and peaked with a significant positive contribution from \textit{Text} channel \casebox{14}. This discovery prompted \textbf{\textit{E1}} to investigate further in the \textit{Performance Inspection} part. 
While analyzing the sales pitch employed predominantly during this time, \textbf{\textit{E1}} discovered that it had transitioned from the \textit{atmosphere pitch} to the \textit{selling pitch} and \textit{order pitch} \casebox{15}. Upon reviewing the corresponding video content \casebox{16}, he noted that the streamer began by providing a historical tidbit relevant to the product to capture the viewers' interest before transitioning to promoting the product. Subsequently, \textbf{\textit{E1}} delved deeper into the \textit{Correlation Modeling} part and meticulously analyzed the \textit{Text} feature, ultimately concluding that all three sales pitches mentioned above had a positive impact on the target data during this period \casebox{17}. Upon comparison with the aforementioned \textit{Drawbacks}, \textbf{\textit{E1}} realized that the optimal employment and arrangement of diverse sales pitches were the primary drivers behind the \textit{GPM} peaks. He noted that the judicious use of the \textit{atmosphere pitch} to attract viewers' attention followed by a seamless transition into \textit{selling pitch} and \textit{order pitch} for promoting the products proved to be a wise choice. Finally, \textbf{\textit{E1}} appended this period to the \textit{Record view} under the category of \textit{Highlights} \casebox{18} for future reference.

\par \textbf{Takeaway message.} By employing predictive modeling and conducting case studies involving streamers, we have discovered that sales pitches exert a significant influence on streaming statistics. This underscores the importance for streamers and researchers to place greater emphasis on sales pitches, particularly their combinations, in order to achieve optimal results and outcomes.

\subsection{Case II: Enlightenment for Marketing Researchers}
\par In Case II, \textbf{\textit{E5}} exhibited significant interests in the viability of conventional marketing and communication theories in this scenario.

\begin{figure}[h]
 \centering 
  \vspace{-3mm}
 \includegraphics[width=\columnwidth]{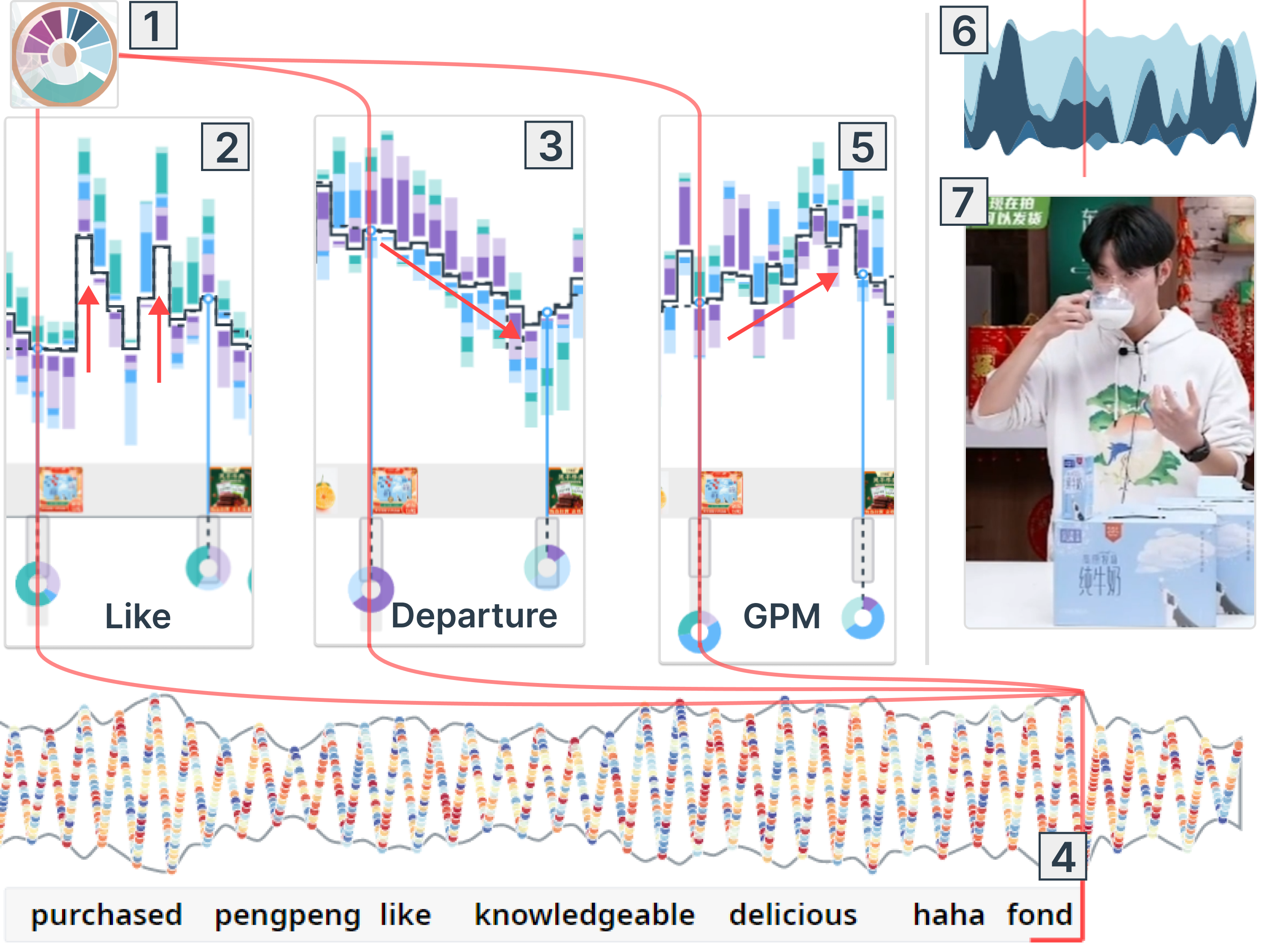}
 \vspace{-6mm}
 \caption{Case II: Enlightenment for Marketing Researchers. \protect\casebox{1} Identifying effective glyphs for combining different sales pitches, \protect\casebox{2} peak in \textit{Like} value during the period, with a positive contribution from the \textit{Text} channel, \protect\casebox{3} declining trend and a negative contribution from the \textit{Text} channel in \textit{Departure}, \protect\casebox{4} observe the keywords like ``purchased, like, delicious, fond''. \protect\casebox{5} Increasing trend in GPM value during the period, with a potential \textit{S-O-R} paradigm, \protect\casebox{6} various sales pitches were used, and \protect\casebox{7} the streamer was demonstrating milk's quality by scooping and savoring it.}
 \label{fig:case2}
  \vspace{-3mm}
\end{figure}

\par \textbf{The \textit{S-O-R} paradigm in livestream e-commerce.} According to Jacoby's ~\cite{jacoby2002stimulus} \textit{Stimulus-Organism-Response} (\textit{S-O-R}) model, external stimuli can affect the cognitive and emotional states of individuals, which in turn can influence their behavioral responses. In the context of livestream e-commerce, \textbf{\textit{E5}} aims to investigate the applicability of the \textit{S-O-R} paradigm. Particularly, \textbf{\textit{E5}} employed the \textit{S-O-R} model as a theoretical framework to investigate the interplay between the streamer's performance, feedback from viewers, and sales data, which were respectively interpreted as the \textit{Stimulus}, \textit{Organism}, and \textit{Response}. Given that the sales pitch is a crucial persuasive factor that directly affects a viewer's purchase decision, \textbf{\textit{E5}} focused on the \textit{Text} channel during the \textit{Stimulus} process. Initially, \textbf{\textit{E5}} analyzed the usage of sales pitch in the \textit{Segment View} and identified glyphs that combined different sales pitches (\cref{fig:case2}-\casebox{1}). To understand the impact of these glyphs on viewer feedback, \textbf{\textit{E5}} examined one of them in detail using the \textit{Exploration View} and used \textit{Like} as the target in the \textit{Correlation Modeling} part. The results showed that the \textit{Like} value peaked during this period, with a positive contribution from the \textit{Text} channel \casebox{2}. To validate this finding, \textbf{\textit{E5}} changed the target to \textit{Departure}, which revealed a declining trend and negative contribution from the \textit{Text} channel \casebox{3}. Additionally, \textbf{\textit{E5}} investigated viewer feedback further by analyzing the \textit{Comment Reference} part and discovered a high intensity of comments containing keywords like ``\textit{purchased, like, delicious, fond}'' \casebox{4}, which suggested positive viewer feedback. Finally, \textbf{\textit{E5}} examined the sales data by changing the target into \textit{GPM} to explore the response. As expected, the \textit{GPM} value demonstrated an increasing trend during this period \casebox{5}. Following the aforementioned exploration process, the feasibility of the \textit{S-O-R} model in the context of livestream e-commerce was ascertained by \textbf{\textit{E5}}. The finding has motivated \textbf{\textit{E5}} to conduct additional research on livestream e-commerce using this model. In general, \textbf{\textit{E5}}'s investigation of the \textit{S-O-R} model has uncovered new possibilities for comprehending the intricate dynamics of livestream e-commerce. This could lead to more efficient and specific strategies in this field.

\par \textbf{Role of streamers from the perspective of McGuire's \textit{Communication Persuasion Matrix.}} McGuire's \textit{Communication Persuasion Matrix}~\cite{mcguire1984public} encompasses input communication variables such as source, message, channel, receiver, and destination, and output mediational steps that underlie the persuasion process in the public communication campaigns. In this regard, \textbf{\textit{E5}} exhibited keen interest in investigating the distinctive source factor, streamers, within the context of marketing communication in streaming media. Building upon previous explorations, \textbf{\textit{E5}} examined the \textit{Performance Inspection part} in the \textit{Exploration View} and observed various sales pitches being used during this period \casebox{6}. However, \textbf{\textit{E5}} recognized the need to delve deeper into the content of these sales pitches to explore the persuasive impact of streamers as communication factors. Consequently, \textbf{\textit{E5}} brushed this period for further exploration of details. By referencing video content, \textbf{\textit{E5}} discovered that the streamer employed narrative and presentation techniques to market milk, such as demonstrating its quality by scooping it and savoring it \casebox{7}. \textbf{\textit{E5}} posited that such a personalized presentation of the product could engender an authentic sense and foster credibility with prospective customers, thereby augmenting the streamer's persuasive effectiveness as a source. Moreover, the streamer customized the explanation of milk's benefits for different age groups, leading \textbf{\textit{E5}} to hypothesize that this approach could enhance viewer engagement and foster a sense of being understood, ultimately cultivating stronger connections with the viewer as a source. By combining the aforementioned peaks in \textit{GPM} and \textit{Like} values within the \textit{Correlation Modeling} part with the corresponding video content, \textbf{\textit{E5}} discerned that the streamer's persuasive approach yielded remarkable outcomes. Moreover, \textbf{\textit{E5}} identified that this process conforms to McGuire's \textit{Communication Persuasion Matrix}, which emphasizes the importance of source characteristics in persuading viewers. \textbf{\textit{E5}} believed that these findings could prove valuable to businesses seeking to employ streaming media as a marketing communication tool, as they offered a framework for crafting compelling and impactful content that resonates with diverse viewers.

\subsection{Expert Interviews}
\par To assess the effectiveness of \textit{LiveRetro} compared to a commercial system previously utilized by streamers (Fig.1 in the appendix), we conducted semi-structured interviews with \textbf{\textit{E1}}-\textbf{\textit{E5}}, which included a 15-minute tutorial and lasted about an hour in total. During the sessions, participants used both systems for approximately 50 minutes to conduct strategic retrospective analyses on randomly selected sessions.

\par \textbf{System Performance and Design.} Experts confirmed \textit{LiveRetro}'s effectiveness in strategic retrospective analysis in livestream e-commerce and correlating video content with synchronous feedback. \textbf{\textit{E1}} and \textbf{\textit{E3}} praised its multi-level contribution summaries for quickly revealing performance features that influence the target. \textbf{\textit{E3}} noted that \textit{LiveRetro} provides a more holistic perspective and accelerates retrospective analysis, while \textbf{\textit{E4}} found it helpful in examining marketing theories in the context of livestream e-commerce. Experts confirmed that the visualizations are valuable, easy to understand, and with fluid interactions. They favored the \textit{Time linkage Mechanism} and the vertically synchronized cursor for their ``\textit{dynamic and immediate access to information}''. The \textit{Exploration View} was also appreciated, with \textbf{\textit{E5}} finding the \textit{Performance Inspection} intuitive and potentially ``\textit{saving weeks of work}''. \textbf{\textit{E1}} and \textbf{\textit{E2}} found the \textit{Correlation Modeling} highly useful for exploring feature contributions from different channels, while \textbf{\textit{E3}} perceived it as complex due to the amount of information visualized.

\par \textbf{Learning Curve.} The experts noted that \textit{LiveRetro}, with more components, took some time (usually a 20-minute trial) to get accustomed to, in contrast to the commercial system. However, they found \textit{LiveRetro} to be incredibly useful for understanding live streaming practices. They gained valuable insights into the relationship between live performance and synchronous feedback and expressed interest in using it for retrospective analysis and theory validation in the future.

\section{Discussion and Limitation}
\noindent\textbf{Lessons Learned.} In our collaboration with experts, we learn three important lessons. First, livestream e-commerce holds significant untapped potential for exploration, encompassing various unexplored scenarios, factors, and relationships. Second, specific guidelines for designing visual interfaces in this domain are essential, as existing literature can result in overwhelming interfaces. Through expert input, we provide tailored principles for future design. Third, achieving ``Alignment'' in joint video content and synchronous feedback analysis is paramount. Our system ensures alignment at multiple levels, enhancing organization and cognitive efficiency.

\noindent\textbf{Practical and Theoretical Inspirations.} Our study reveals that family life segments in live streaming have a higher average viewer engagement time compared to informative segments. This finding leads our streamers to focus on emotional resonance to increase viewer retention, and marketing researchers appreciate our categorization of sales pitches. Our work encourages further research on the effects of different sales pitches and moves beyond general streamer characteristics.

\noindent\textbf{Generalizability and Scalability.} \textit{LiveRetro} enables comprehensive and strategic retrospective analysis of livestream e-commerce and has been successfully tested on \textit{TikTok} with streamers and marketing researchers. It can also be applied to other platforms like \textit{CommentSold}, \textit{Bambuser}, and \textit{Instagram Live} for retrospective analysis. The system can be extended to support joint analysis of video and synchronous feedback in different domains. However, scalability issues arise with feature extraction and visual designs. To address this, techniques such as GPU acceleration and parallel computing can be utilized for efficient facial expression extraction. Moreover, visual clutter in the \textit{Segment View} can be reduced by implementing semantic zooming and click-to-focus features for specific segments of interest.

\noindent\textbf{Limitations.} Due to data regulations, streaming statistics on \textit{TikTok} have a maximum sampling rate of one minute, necessitating the adoption of the same granularity for performance features. However, finer granularity, like two-second intervals, could improve real-time feedback and capture intricate correlations. The current modeling approach lacks consideration of merchandise-specific characteristics, such as category, popularity, and cost-effectiveness, which can influence feedback. Additionally, case studies involved our collaborative team, necessitating further testing to strengthen the system's usability and effectiveness.

\section{Conclusion and Future Work}
\par This study develops design principles and a visual analytics system called \textit{LiveRetro} for retrospective analysis of livestream e-commerce, incorporating computational features and feedback from streamers, viewers, and merchandise. \textit{LiveRetro} features enhanced visualization and multiple time-series forecasting models to model specific targets and identify performance influences at different levels. Two case studies and expert interviews demonstrate \textit{LiveRetro}'s ability to provide deep insights and enable efficient retrospective analysis. Future plans include incorporating merchandise factors and providing comparative analysis at market and account levels.

\acknowledgments{This work is partially supported by the Shanghai Frontiers Science Center of Human-centered Artificial Intelligence (ShangHAI) and Key Laboratory of Intelligent Perception and Human-Machine Collaboration (ShanghaiTech University), Ministry of Education.}

\balance
\bibliographystyle{abbrv-doi-hyperref-narrow}

\bibliography{template}








\end{document}